\documentclass{article}

\usepackage{PRIMEarxiv}

\usepackage[utf8]{inputenc} 
\usepackage[T1]{fontenc}    
\usepackage{hyperref}       
\usepackage{url}            
\usepackage{booktabs}       
\usepackage{amsfonts}       
\usepackage{nicefrac}       
\usepackage{microtype}      
\usepackage{lipsum}
\usepackage{fancyhdr}       
\usepackage{graphicx}       
\graphicspath{{media/}}     
\usepackage{xcolor}
\hypersetup{colorlinks=true, linkcolor=blue, filecolor=black, urlcolor=blue, citecolor=cyan}
\usepackage[acronym,toc]{glossaries}
\usepackage[intoc]{nomencl}
\usepackage{appendix}

\usepackage{amsfonts, amsmath, amsthm, amssymb}
\usepackage{newtxtext,newtxmath}
\title{Classical Soft Graviton Theorem due to Scalar Fields on 4-D Minkowski Background
}

\author{
  Raikhik Das \\
  Department of Physics \\
  Indian Institute of Science Education and Research, Pune \\
  Dr. Homi Bhabha Rd, Pashan, Pune, Maharashtra 411008, India\\
  \texttt{raikhikd@gmail.com} \\
}

\numberwithin{equation}{section}

\begin{document}
\maketitle

\begin{abstract}
The classical soft graviton theorem expresses the behavior of low-frequency gravitational radiation. In this paper, simplistic proofs of the classical soft graviton theorem for massless and massive scalar fields on $4$-D Minkowski background are presented without considering the correction to the behavior of scalar fields and gravitational stress-energy tensor due to the perturbation in the background.
\end{abstract}


\section*{\centering Conventions and Symbols}
\begin{center}

$\begin{array}{lll}
     \mathbf{Re}[z] \equiv \text{ Real part of complex number } z & \hspace{1cm} & \mathcal{I}^+ : \text{ Future Null Infinity}\\
     \mathbf{Im}[z] \equiv \text{ Imaginary part of complex number } z & \hspace{1cm} & \mathcal{I}^+_\pm : u \to \pm \infty \text{ at Future Null Infinity} \\
     \text{Signature of metric: } (-,+,+,+) &\hspace{1cm}  & K_{(\mu} L_{\nu)} = K_{\mu} L_{\nu} + K_{\nu} L_{\mu}  
\end{array}$
    
\end{center}

\section{Introduction}

The spectrum of low-frequency gravitational radiation emitted during a classical scattering process is characterized by the classical soft graviton theorem \cite{Sen, Sen_Laddha, Sahoo2022}. This work aims to demonstrate that general relativity coupled to matter equations of motion can be directly used to derive the classical soft graviton theorem. One can further investigate the effects of perturbation to the background metric on the matter system's behavior. The perturbation to the background metric will modify the behavior of the matter system, leading to the correction of the behavior of the soft radiation \cite{Sen}. The gravitational stress-energy tensor arising due to the perturbation to the background metric will also result in the correction of the behavior of soft radiation. Upon considering the effects of perturbation, logarithmic corrections are introduced to the soft expansion of gravitational radiation. Considering the effects of perturbation to the background metric arising in the behavior of gravitational radiation at soft limit will lead to sub-leading and further higher order corrections. But in this article, the effects of perturbation to the metric on the whole system are not considered. Before delving into the proof of the classical soft graviton theorem, a brief review of linearized gravitational radiation has been presented in sec.[\ref{Sec_Lin_Grav_Rad}]. Then, the simplistic proofs of the classical soft graviton theorem for real massless scalar field and real massive scalar field have been demonstrated in sec.[\ref{Sec_Soft Gravitional Radiation for Massless Scalar Fields}] and sec.[\ref{Sec_Soft Gravitons for Massive Scalar Field}], respectively. 


\section{Linearized Gravitational Radiation}\label{Sec_Lin_Grav_Rad}

To study gravitational radiation, one can study linearized perturbation to the Minkowski background. So, let's assume that there exists a metric $g_{\mu\nu}$ such that
\begin{equation}
    g_{\mu\nu} = \eta_{\mu\nu} + 2 h_{\mu\nu},
    \label{GW_1}
\end{equation}
where $\eta_{\mu\nu}$ is the Minkowski metric and $h_{\mu\nu}$ part represents the perturbation to $\eta_{\mu\nu}$.\\

Now, let's define:
\begin{equation}
    e_{\mu\nu} = h_{\mu\nu} - \frac{1}{2} \eta_{\mu\nu} \eta^{\alpha\beta} h_{\alpha\beta}
    \label{GW_2}
\end{equation}

    \label{GW_3}

In de-Donder coordinates or harmonic coordinates, Einstein's equations of general relativity for $g_{\mu\nu}$ reduces to:
\begin{equation}
    \square e_{\mu\nu} = -8 \pi G T_{\mu\nu},
    \label{GW_4}
\end{equation}
where $\square \equiv \eta_{\alpha\beta \partial_\alpha \partial_\beta}$, $G$ is Newton's universal constant and $T_{\mu\nu}$ is the stress-energy tensor.\\

For convenience, this article works in units for which $8 \pi G = 1$. Therefore, eq.\eqref{GW_4} takes the form:
\begin{equation}
    \square e_{\mu\nu}(x) = - T_{\mu\nu}(x).
    \label{GW_5}
\end{equation}

The retarded solution of eq.\eqref{GW_5} is given in eq.\eqref{Rad_fld_8}.

\begin{equation}
    e_{\mu\nu}(x) = -\int d^4 x^\prime  G_r (x, x^\prime) T_{\mu\nu}(x),
    \label{Rad_fld_8}
\end{equation}
where $G_r (x, x^\prime)$ represents the retarded Green's function:
\begin{equation}
    G_r (x, x^\prime) = \int \frac{dl^4}{(2 \pi)^4} \frac{e^{i l \cdot (x - x^\prime)}}{(l_0 + i \epsilon)^2 - \vec{l}^2}.
    \label{Rad_fld_9}
\end{equation}\\

Now, upon Fourier transforming ${e}_{\mu\nu}(x)$, one gets
\begin{align}
    \Tilde{e}_{\mu\nu}(\omega, \vec{x}) \nonumber &=  -\int dt e^{ i \omega t} \int d^4 x^\prime  \int \frac{dl^4}{(2 \pi)^4} \frac{e^{i l \cdot (x - x^\prime)}}{(l_0 + i \epsilon)^2 - \vec{l}^2} T_{\mu\nu}(x^\prime) \\ \nonumber&
     = -\int d^4 x^\prime  \int \frac{dl^4}{(2 \pi)^3} \frac{e^{i l_0 x^{\prime 0} +i\Vec{l}\cdot (\vec{x} - \vec{x^\prime})}}{(l_0 + i \epsilon)^2 - \vec{l}^2} T_{\mu\nu}(x^\prime) \int \frac{dt}{2 \pi} e^{ i (\omega-l_0) t} \\ 
    \nonumber& = -\int d^4 x^\prime  \int \frac{dl^3}{(2 \pi)^3} \frac{e^{i \omega x^{\prime 0} +i\Vec{l}\cdot (\vec{x} - \vec{x^\prime})}}{(\omega + i \epsilon)^2 - \vec{l}^2} T_{\mu\nu}(x^\prime) \\ 
    \nonumber& =  -\int d^4 x^\prime  \int \frac{dl_\perp^2}{(2 \pi)^2} \frac{dl_\parallel}{(2 \pi)} \frac{e^{i \omega x^{\prime 0} +i l_\parallel |\vec{x} - \vec{x^\prime}|}}{(\omega + i \epsilon)^2 - \vec{l}^2_\perp - l_\parallel^2} T_{\mu\nu}(x^\prime) \\ 
    \nonumber& =  i\int d^4 x^\prime  \int \frac{dl_\perp^2}{(2 \pi)^2} \frac{e^{i \omega x^{\prime 0} +i\sqrt{(\omega + i \epsilon)^2 - \vec{l}^2_\perp} |(\vec{x} - \vec{x^\prime})|}}{2\sqrt{(\omega + i \epsilon)^2 - \vec{l}^2_\perp}} T_{\mu\nu}(x^\prime) \\ 
    \nonumber& =  i\int d^4 x^\prime \frac{e^{i \omega x^{\prime 0} +i (\omega + i \epsilon) |(\vec{x} - \vec{x^\prime})|}}{2  (\omega + i \epsilon)} \frac{ (\omega + i \epsilon)}{2 \pi i |(\vec{x} - \vec{x^\prime})|} T_{\mu\nu}(x^\prime) \\ 
    &\simeq \frac{e^{i \omega |\Vec{x}|}}{4 \pi |\vec{x}|}\int d^4 x^\prime e^{-i k \cdot x^\prime} T_{\mu\nu}(x^\prime),
    \label{Rad_fld_10}
\end{align}

In eq.\eqref{Rad_fld_10}, it has been assumed that $|\Vec{x}| >> |\Vec{x^\prime}|$.\\

Using the expression: $\int \frac{dt}{2 \pi} e^{ i (\omega-l_0) t} = \delta(l_0 - \omega)$, one can proceed from the second step of eq.\eqref{Rad_fld_10} to the third step. $\Vec{l}$ is separated into components parallel and perpendicular to $(\vec{x} - \vec{x^\prime})$, referred to as $l_\parallel$ and $\Vec{l}_\perp$, respectively, in the fourth step. Hence, one can observe that $(\omega + i \epsilon)^2 - \vec{l}^2_\perp - l_\parallel^2=0 $ has two roots: $\pm \sqrt{(\omega + i \epsilon)^2 - \vec{l}^2_\perp}$. By doing the contour integration, the fifth step is reached. In the fifth step, the exponent is a rapidly fluctuating function of $\Vec{l}_\perp$ for large $|(\vec{x} - \vec{x^\prime})|$. Therefore, one can integrate over $\Vec{l}_\perp$ using saddle point approximation.\\

One can write, 
\begin{equation}
     \Tilde{e}_{\mu\nu}(\omega, \vec{x}) = \frac{e^{i \omega r}}{4 \pi r}\int d^4 x^\prime e^{-i k \cdot x^\prime} T_{\mu\nu}(x^\prime),
    \label{GW_6}
\end{equation}
where $r=|\Vec{x}|$, $n_i = \frac{x_i}{r}$ and $k \equiv \omega (1, \Vec{n})$.

\section{Soft Gravitional Radiation for Massless Scalar Fields}\label{Sec_Soft Gravitional Radiation for Massless Scalar Fields}

This section provides the derivation of the classical soft graviton theorem for real massless scalar field on $4$-D Minkowski background. The corrections in the behavior of the massless scalar field and the gravitational stress energy tensor due to the perturbation in the background will not be considered. The stress-energy tensor will be used upto the order: $\mathcal{O}\left( \frac{1}{r^3} \right)$. This will help calculate the behavior of gravitational radiation at soft limit upto the  order: $\mathcal{O}(1)$.

\subsection{Preliminaries}\label{Sec_Soft Gravitons for Massless Scalar Field_1}

The data on outgoing null geodesics can describe the massless scalar field space. The chosen coordinates for this description are the retarded null coordinates, where $ u = t-r$. In the retarded null coordinates, the line element takes the form as follows:
\begin{equation}
    ds^2=-du^2 -2 du dr + r^2 d\Omega^2,
    \label{mlSclr_1}
\end{equation}
where $d\Omega^2$ is the unit sphere metric. 

This analysis is being done in de-Donder coordinates. So, let's first define:
\begin{equation}
    \partial_\mu r = \tilde{n}_\mu \hspace{3mm},\hspace{3mm} \partial_\mu u = -n_\mu,\hspace{3mm} \partial_\mu \hat{n}_i = \frac{1}{r} \eta^\perp_{\mu i} ,
    \label{mlSclr_2}
\end{equation}
where $\eta^\perp_{\mu \nu} = \eta_{\mu \nu } + n_\mu n_\nu - n_\mu \tilde{n}_\nu - n_\nu \tilde{n}_\mu$ with $\hat{n}_i=\frac{x_i}{r}$, $n=(1, \vec{\hat{n}})$ and $\tilde{n}=(0, \vec{\hat{n}})$.\\

For a general function $\mathcal{F}(u, r, \hat{n})$,
\begin{align}
    \partial_\mu \mathcal{F}(u, r, \hat{n})  \noindent\nonumber&= \partial_\mu u \frac{\partial \mathcal{F}}{\partial u} + \partial_\mu  r\frac{\partial \mathcal{F}}{\partial r} + \frac{1}{r} \eta^\perp_{\mu i} \partial_\perp^i \mathcal{F} \\
     \noindent& =-n_\mu \frac{\partial \mathcal{F}}{\partial u} + \tilde{n}_\mu \frac{\partial \mathcal{F}}{\partial r} + \frac{1}{r} \eta^\perp_{\mu i} \partial_\perp^i \mathcal{F}
    \label{mlSclr_3}
\end{align}

\subsection{Proof of Classical Soft Graviton Theorem for Massless Scalar Field on 4-D Minkowski Background}\label{Sec_Soft Gravitons for Massless Scalar Field_2}

In this subsection, $\eta_{\mu\nu}$ is considered to be the background metric. The solution to $ g^{\mu\nu} D_\mu D_\nu \, \varphi(x) = 0$, can be expanded near future null infinity in the following manner:

\begin{equation}
    \varphi (u, r, \hat{n}) = \frac{\varphi^{(1)}(u, \hat{n})}{r}  + \frac{\varphi^{(2)}(u, \hat{n})}{r^2} + \ldots.
    \label{mlSclr_4}
\end{equation}

The stress-energy tensor for a massless scalar field is
\begin{equation}
    T^{(\varphi)}_{\mu\nu} (x) = (\partial_\mu \varphi \partial_\nu \varphi) - \eta_{\mu \nu} \mathcal{L}^{(\varphi)},
    \label{mlSclr_5}
\end{equation}
where the Lagrangian of the massless scalar field is $\mathcal{L}=\frac{1}{2}\partial_\mu \varphi \partial^\mu \varphi$.\\

The components of the stress-energy tensor for the massless scalar field as $r \to \infty$ take the following form:
\begin{equation}
    T^{(\varphi)R}_{\mu\nu} (x) = \frac{A_{\mu\nu}(u, \hat{n})}{r^2} +  \frac{B_{\mu\nu}(u, \hat{n})}{r^3}+ \mathcal{O}(\frac{1}{r^4}),
    \label{mlSclr_6}
\end{equation}
where$ \begin{array}[t]{lcl}
    A_{\mu\nu}(u, \hat{n})=\partial_u \varphi^{(1)}(u, \hat{n})\partial_u \varphi^{(1)}(u, \hat{n}) n_\mu n_\nu \\ 
    \\
    B_{\mu\nu}(u, \hat{n})=\begin{array}[t]{lcl}
    \varphi^{(1)}(u, \hat{n})\partial_u \, \varphi^{(1)}(u, \hat{n})n_{(\mu} \tilde{n}_{\nu)} \,-\, \eta^\perp_{(\mu i}\partial^i_\perp \varphi^{(1)}(u, \hat{n})\partial_u \, \varphi^{(1)}(u, \hat{n})n_{\nu)} \,+\, 2 \partial_u \varphi^{(1)}(u, \hat{n}) \partial_u \varphi^{(2)}(u, \hat{n})n_\mu n_\nu\\
    -\, \eta_{\mu\nu} \varphi^{(1)}(u, \hat{n})\partial_u \, \varphi^{(1)}(u, \hat{n}) 
    \end{array}
\end{array}
$

To express $B_{\mu\nu}(u, \hat{n})$ in terms of only the free data $\varphi^{(1)}(u,\hat{n})$, one can use the equation of motion for the massless scalar field and get $\partial_u \varphi^{(2)}(u, \hat{n}) = - \frac{1}{2} \partial_\perp^j \eta^\perp_{jk} \partial_\perp^k \varphi^{(1)}(u,\hat{n})$. Therefore, in terms of free data $\varphi^{(1)}(u,\hat{n})$,

$B_{\mu\nu}(u, \hat{n})=\begin{array}[t]{lcl}
    \varphi^{(1)}(u, \hat{n})\partial_u \, \varphi^{(1)}(u, \hat{n})n_{(\mu} \tilde{n}_{\nu)} \,-\, \eta^\perp_{(\mu i}\partial^i_\perp \varphi^{(1)}(u, \hat{n})\partial_u \, \varphi^{(1)}(u, \hat{n})n_{\nu)} \,-\,  \partial_u \varphi^{(1)}(u, \hat{n}) \partial_\perp^j \eta^\perp_{jk} \partial_\perp^k \varphi^{(1)}(u,\hat{n}) n_\mu n_\nu\\
    -\, \eta_{\mu\nu} \varphi^{(1)}(u, \hat{n})\partial_u \, \varphi^{(1)}(u, \hat{n}) 
    \end{array}$.\\

Following eq.\eqref{GW_6}, one can write:
\begin{equation}
    \Tilde{e}_{\mu\nu} (\omega, \Vec{x})=\frac{e^{i\omega r}}{4 \pi r}\int dx^{\prime 4} e^{-i k \cdot x^\prime}  T^{(\varphi)}_{\mu\nu}=\Tilde{e}^1_{\mu\nu} (\omega, \Vec{x})+\Tilde{e}^2_{\mu\nu} (\omega, \Vec{x})
    \label{mlSclr_7},
\end{equation}
where $$\Tilde{e}^1_{\mu\nu} (\omega, \Vec{x})= \frac{e^{i\omega r}}{4 \pi r}\int_{r>r_0} dx^{\prime 4} \, e^{-i k \cdot x^\prime} \, T^{(\varphi)R}_{\mu\nu} (x^\prime)$$ and $$\Tilde{e}^2_{\mu\nu} (\omega, \Vec{x})= \frac{e^{i\omega r}}{4 \pi r}\int_{r\leq r_0} dx^{\prime 4} \, e^{-i k \cdot x^\prime} \, T^{(\varphi)}_{\mu\nu} (x^\prime)$$.\\

Using the asymptotic form of the stress-energy tensor for the massless scalar field eq.\eqref{mlSclr_6} in $\Tilde{e}^1_{\mu\nu} (\omega, \Vec{x})$, one gets
\begin{align}
    \Tilde{e}^1_{\mu\nu} (\omega, \Vec{x}) \noindent\nonumber&=  \frac{e^{i\omega r}}{4 \pi r}\int dx^{\prime 4} \, e^{-i k \cdot x^\prime} \Bigg( \frac{A_{\mu\nu}(u^\prime, \hat{n}^\prime)}{r^{\prime 2}} +  \frac{B_{\mu\nu}(u^\prime, \hat{n}^\prime)}{r^{\prime 3}}  + \mathcal{O}\left(\frac{1}{r^{\prime 4}}\right)\Bigg)  \\ 
     \noindent\nonumber&=  \frac{e^{i\omega r}}{4 \pi r}\int r^2 du^\prime dr^\prime d\hat{n}^\prime e^{i \omega u^\prime + i \omega r^\prime - i \omega \Vec{n}\cdot \Vec{n}^\prime r^\prime} \Bigg( \frac{A_{\mu\nu}(u^\prime, \hat{n}^\prime)}{r^{\prime 2}} +  \frac{B_{\mu\nu}(u^\prime, \hat{n}^\prime)}{r^{\prime 3}}  + \mathcal{O}\left(\frac{1}{r^{\prime 4}}\right)\Bigg)\\
      \noindent\nonumber&=  \frac{e^{i\omega r}}{4 \pi r}\int du^\prime dr^\prime d\hat{n}^\prime e^{i \omega u^\prime + i \omega r^\prime - i \omega \Vec{n}\cdot \Vec{n}^\prime r^\prime} \Bigg( A_{\mu\nu}(u^\prime, \hat{n}^\prime) +  \frac{B_{\mu\nu}(u^\prime, \hat{n}^\prime)}{r^{\prime}}  + \mathcal{O}\left(\frac{1}{r^{\prime 4}}\right)\Bigg) \\
      \noindent\nonumber&=   \frac{e^{i\omega r}}{4 \pi r}\int du^\prime dr^\prime d\hat{n}^\prime e^{i \omega u^\prime + i \omega r^\prime - i \omega \Vec{n}\cdot \Vec{n}^\prime r^\prime}{A_{\mu\nu}(u^\prime, \hat{n}^\prime)} \\
     \noindent\nonumber&  \hspace{5mm}+\, \frac{e^{i\omega r}}{4 \pi r}\int du^\prime dr^\prime d\hat{n}^\prime e^{i \omega u^\prime + i \omega r^\prime - i \omega \Vec{n}\cdot \Vec{n}^\prime r^\prime}\frac{B_{\mu\nu}(u^\prime, \hat{n}^\prime)}{r^{\prime }}  \\
     \noindent& \hspace{10mm}+\, \frac{e^{i\omega r}}{4 \pi r}\int du^\prime dr^\prime d\hat{n}^\prime e^{i \omega u^\prime + i \omega r^\prime - i \omega \Vec{n}\cdot \Vec{n}^\prime r^\prime}\left(\mathcal{O}\left(\frac{1}{r^{\prime 2}}\right)\right)
     \label{mlSclr_8}
\end{align}

The following relations have been used in the second step of eq.\eqref{mlSclr_8}:
\begin{equation}
    dx^{\prime 4} \to r^{\prime 2} du^\prime dr^\prime d\hat{n}^\prime \hspace{3mm},\hspace{3mm} k \cdot x^\prime = -(\omega \, u^\prime +  \omega r^\prime -  \omega \, \Vec{n}\cdot \Vec{n}^\prime r^\prime)=-\omega \, u^\prime + \omega \, n \cdot n^\prime r^\prime,
    \label{mlSclr_9}
\end{equation}
where $k^\mu \equiv \omega(1, \vec{n})$.\\

Now, let's calculate the first term in the last step of eq.\eqref{mlSclr_8}:
\begin{align}
    \noindent\nonumber&\hspace{2mm} \frac{e^{i\omega r}}{4 \pi r}\int du^\prime d\hat{n}^\prime \int_{r_0}^\infty dr^\prime e^{i \omega u^\prime - i \omega \, n \cdot n^\prime r^\prime}A_{\mu\nu}(u^\prime, \hat{n}^\prime) \\
    \sim \noindent\nonumber&\hspace{2mm} \frac{e^{i\omega r}}{4 \pi r}\int du^\prime d\hat{n}^\prime  \frac{e^{i \omega u^\prime - i \omega \, n \cdot n^\prime r_0}}{i \omega \, n \cdot n^\prime}A_{\mu\nu}(u^\prime, \hat{n}^\prime) \\
    = \noindent&\hspace{2mm} \frac{e^{i\omega r}}{4 \pi r}\frac{1}{\omega}\int du^\prime d\hat{n}^\prime  \frac{1}{i \omega \, n \cdot n^\prime}A_{\mu\nu}(u^\prime, \hat{n}^\prime) + \mathcal{O}(1)
    \label{mlSclr_10}.
\end{align}
In the second step of the eq.\eqref{mlSclr_10}, the terms at infinity we has 
 been ignored.\\

Now let's calculate the second term in the last step of eq.\eqref{mlSclr_8}.

\begin{align}
    \noindent\nonumber&\hspace{2mm}\frac{e^{i\omega r}}{4 \pi r}\int du^\prime d\hat{n}^\prime \int_{r_0}^\infty dr^\prime e^{i \omega u^\prime - i \omega \, n \cdot n^\prime r^\prime}\frac{B_{\mu\nu}(u^\prime, \hat{n}^\prime)}{r^{\prime }} \\
    = \noindent&\hspace{2mm} -\frac{e^{i\omega r}}{4 \pi r} \, \log{\omega}\int du^\prime d\hat{n}^\prime B_{\mu\nu}(u^\prime, \hat{n}^\prime) + \mathcal{O}(1)
     \label{mlSclr_11}
\end{align}

The third term in the last step of eq.\eqref{mlSclr_8} is of no interest here as it will result in at least $\mathcal{O}(1)$ terms.\\

Now, let's calculate $\Tilde{e}^2_{\mu\nu} (\omega, \Vec{x})$. To do so, one can write
\begin{align}
    k_\alpha \Tilde{e}^{2\alpha \beta} (\omega, \Vec{x})  \noindent\nonumber&= i\frac{e^{i\omega r}}{4 \pi r}\int_{r\leq r_0} dx^{\prime 4} \Big{\{} \frac{\partial}{\partial x^{\prime \alpha }}e^{-i k \cdot x^\prime}\Big{\}} T^{(\varphi) \alpha \beta} (x^\prime) \\
    \noindent&= i\frac{e^{i\omega r}}{4 \pi r}\int d \hat{n}^{\prime} \int d u^{\prime }r^{\prime 2} \hat{n}_{\alpha}^{\prime} e^{-i k \cdot x^{\prime}} T^{(\varphi)R \alpha \beta}(x^{\prime})\big|_{r^{\prime}=r_0},
    \label{mlSclr_12}
\end{align}
where integration by parts was done in the second step. The boundary term has been picked up at $r^{\prime}=r_0$ and used the conservation law: $\partial_{\alpha} T^{(\varphi)R\alpha \beta}\left(x^{\prime}\right)=0$.

The sum of the incoming flux and outgoing momentum flux is equal. Therefore,
\begin{equation}
    r^{\prime 2} \int d \hat{n}^{\prime} \int d u^{\prime} \hat{n}_{\alpha}^{\prime} T^{(\varphi)R\alpha \beta}x^{\prime}\big|_{r^{\prime}=r_0}=0.
    \label{mlSclr_13}
\end{equation}

Using eq.\eqref{mlSclr_13} in eq.\eqref{mlSclr_12}, one gets
\begin{equation}
    k_{\alpha} \tilde{e}^{2 \alpha \beta}(x)=\left.\frac{e^{i\omega r}}{4 \pi r} r^{\prime 2} \int d \hat{n}^{\prime} \int d u^{\prime} \hat{n}_{\alpha}^{\prime} k \cdot x^{\prime} T^{(\varphi)R \alpha \beta}\left(x^{\prime}\right)\right|_{r^{\prime}=r_0}+\mathcal{O}\left(\omega^{2}\right).
    \label{mlSclr_14}
\end{equation}

One can take the solution of eq.\eqref{mlSclr_14} to be
\begin{equation}
    \tilde{e}^{2 \alpha \beta}(x)=\left. \frac{e^{i\omega r}}{4 \pi r} r^{\prime 2} \int d \hat{n}^{\prime} \int d u^{\prime} \hat{n}_{\gamma}^{\prime} x^{\prime \alpha} T^{(\varphi)R\gamma \beta}\left(x^{\prime}\right)\right|_{r^{\prime}=r_0}+\mathcal{O}(\omega) .
    \label{mlSclr_15}
\end{equation}
But $\tilde{e}^{2 \alpha \beta}(x)$ in eq.\eqref{mlSclr_15} is not symmetric. To symmetrize $\tilde{e}^{2 \alpha \beta}(x)$, the angular momentum conservation can be used:
\begin{equation}
    \left.r^{\prime 2} \int d \hat{n}^{\prime} \int d u^{\prime} \hat{n}_{\gamma}^{\prime}\left[-x^{\prime \alpha} T^{(\varphi)R \gamma \beta}\left(x^{\prime}\right)+x^{\prime \beta} T^{(\varphi)R \gamma \alpha}\left(x^{\prime}\right)\right]\right|_{r^{\prime}=r_0}=0.
    \label{mlSclr_16}
\end{equation}

The symmetrized $\tilde{e}^{2 \alpha \beta}(x)$ is as following:
\begin{equation}
    \tilde{e}^{2 \alpha \beta}(\omega, \vec{x})=\left.\frac{1}{2} \frac{e^{i\omega r}}{4 \pi r} \int d \hat{n}^{\prime} \int d u^{\prime}r^{\prime 2} \hat{n}_{\gamma}^{\prime}\left[x^{\prime \alpha} T^{(\varphi)R\gamma \beta}\left(x^{\prime}\right)+x^{\prime \beta} T^{(\varphi)R \gamma \alpha}\left(x^{\prime}\right)\right]\right|_{r^{\prime}=r_0}+\mathcal{O}(\omega) .
    \label{mlSclr_17}
\end{equation}
In eq.\eqref{mlSclr_17}, one can see that $\tilde{e}^{2 \alpha \beta}(x)=\mathcal{O}(1)$.\\

Combining eq.\eqref{mlSclr_10}, eq.\eqref{mlSclr_11} and eq,\eqref{mlSclr_17}, one gets the following soft expansion of $\tilde{e}_{\mu \nu}(\omega, \vec{x})$:

\begin{equation}
    \tilde{e}_{\mu \nu}(\omega, \vec{x}) = \frac{e^{i\omega r}}{4 \pi r}\left\{\frac{1}{i \,\omega}\int du^\prime d\hat{n}^\prime  \frac{A_{\mu\nu}(u^\prime, \hat{n}^\prime)}{ n \cdot n^\prime} - \log{\omega}\int du^\prime d\hat{n}^\prime B_{\mu\nu}(u^\prime, \hat{n}^\prime) + \mathcal{O}(1) \right\}.
    \label{mlSclr_18}
\end{equation}

\section{Soft Gravitional Radiation for Massive Scalar Fields}\label{Sec_Soft Gravitons for Massive Scalar Field}
This section provides the derivation of the classical soft graviton theorem for real massive scalar field on $4$-D Minkowski background. The corrections in the behavior of the massless scalar field and the gravitational stress energy tensor due to the perturbation in the background will not be considered here. The stress-energy tensor will be used upto the order: $\mathcal{O}\left( \frac{1}{\tau^5} \right)$. This will help us calculate the behavior of graviational radiation at soft limit upto the order: $\mathcal{O}(1)$. This section will attempt to investigate the soft gravitational radiation at $\mathcal{I}^\pm$.

\subsection{Preliminaries}\label{Sec_Soft Gravitons for Massive Scalar Field_1}

The data on a unit hyperboloid describing timelike infinity can describe the massive scalar field space. The chosen coordinates for this description are
\begin{equation}
    \tau = \pm \sqrt{t^2-r^2}\hspace{3mm},\hspace{3mm} \rho=\frac{r}{\tau}.
    \label{MSclr_1}
\end{equation}

In the new coordinates, the line element takes the form as follows:
\begin{equation}
    ds^2=-d\tau^2 + \frac{\tau^2}{1+\rho^2} d\rho^2 + \rho^2\tau^2 d\Omega^2,
    \label{MSclr_2}
\end{equation}
where $d\Omega^2$ is the unit sphere metric. \\ 

We will treat all the tensors in de-Donder coordinates. So, let us first define:
\begin{equation}
    \mathcal{N}_\mu = \partial_\mu \tau = (\sqrt{1+\rho^2}, - \rho \vec{n}) \hspace{3mm},\hspace{3mm} \mathcal{M}_\mu = \tau \, \partial_\mu \rho = \sqrt{1+\rho^2} (-\rho, \sqrt{1+\rho^2} \, \vec{n}) \hspace{3mm},\hspace{3mm} \partial_\mu n_i = \frac{1}{\rho \tau} \eta^\perp_{\mu i},
    \label{MSclr_3}
\end{equation}
where $\eta^\perp_{\mu \nu} = \eta_{\mu \nu } + n_\mu n_\nu - n_\mu \tilde{n}_\nu - n_\nu \tilde{n}_\mu$ with $\hat{n}_i=\frac{x_i}{r}$, $n=(1, \vec{\hat{n}})$ and $\tilde{n}=(0, \vec{\hat{n}})$.\\

For a general function $\mathcal{G}(\tau, \rho, \hat{n})$,
\begin{equation}
    \partial_\mu \mathcal{G}(\tau, \rho, \hat{n}) \begin{array}[t]{ll}
         = \partial_\mu \tau \frac{\partial \mathcal{G}}{\partial \tau} + \partial_\mu  \rho\frac{\partial \mathcal{G}}{\partial \rho} + \frac{1}{\tau \rho} \eta^\perp_{\mu i} \partial_\perp^i \mathcal{G}  \vspace{3mm}\\ 
         = \mathcal{N}_\mu \frac{\partial \mathcal{G}}{\partial \tau} + \mathcal{M}_\mu \frac{\partial \mathcal{G}}{\partial \rho} + \frac{1}{\tau \rho} \eta^\perp_{\mu i} \partial_\perp^i \mathcal{G}. 
    \end{array}
    \label{MSclr_4}
\end{equation}

\subsection{Proof of Classical Soft Graviton Theorem for Massive Scalar Fields}\label{Sec_Sft_Grav_mSclr}

As $\tau \to \pm \infty$, the massive scalar field behaves as following:
\begin{equation}
    \Phi (\tau, \rho, \hat{n}) = \left(\frac{\Phi^{(3/2)}(\rho, \hat{n})}{|\tau|^{3/2}} + \frac{\Phi^{(5/2)}_\pm(\rho, \hat{n})}{|\tau|^{5/2}}+ \mathcal{O}\left(\frac{1}{|\tau|^{7/2}}\right)\right) e^{-i m \tau} + \left(\frac{\Phi^{(3/2) *}(\rho, \hat{n})}{|\tau|^{3/2}} + \frac{\Phi^{(5/2) *}_\pm(\rho, \hat{n})}{|\tau|^{5/2}} + \mathcal{O}\left(\frac{1}{|\tau|^{7/2}}\right)\right) e^{i m \tau}
    \label{MSclr_5}
\end{equation}

One can get the asymptotic behavior of the massive scalar field in eq.\eqref{MSclr_5} from the equation of motion: $D^\mu D_\mu \Phi (\tau, \rho, \hat{n}) = 0$.

The stress-energy tensor for a massive scalar field is
\begin{equation}
    T^{(\Phi)}_{\mu\nu} (x) = (\partial_\mu \Phi \partial_\nu \Phi) - \eta_{\mu \nu} \mathcal{L}^{(\Phi)} \: , \: \text{where } \mathcal{L}^{(\Phi)} = \frac{1}{2} \partial_\mu \Phi \partial^\mu \Phi + \frac{1}{2} m^2 \Phi^2 .
    \label{MSclr_6}
\end{equation}

The components of stress-energy tensor for the massive scalar field as $\tau \to \pm\infty$ takes the following form\footnote{It will be assumed that this behavior will hold good for $|\tau| > \tau_0$.}:
\begin{equation}
    T^{(\Phi)R}_{\mu\nu} (x) =   \frac{F_{\mu\nu}^\pm(\rho, \hat{n})}{\tau^3} +  \frac{H_{\mu\nu}^\pm(\rho, \hat{n})}{\tau^4}  +\mathcal{O}\bigg(\frac{1}{\tau^5}\bigg),
    \label{MSclr_7}
\end{equation}
where $\begin{array}[t]{lcl}
    F^\pm_{\mu\nu} (\rho, \hat{n}) = \pm 2 m^2 \, \Phi^{(3/2)}(\rho, \hat{n}) \Phi^{*(3/2)} (\rho, \hat{n}) \mathcal{N}_\mu \mathcal{N}_\nu \\ 
    \\
    H^\pm_{\mu\nu} (\rho, \hat{n}) =\begin{array}[t]{lcl}
    4 m^2 \, \mathbf{Re}[\Phi^{(3/2)}(\rho, \hat{n}) \Phi^{*(5/2)} (\rho, \hat{n})] \mathcal{N}_\mu \mathcal{N}_\nu \mp \frac{2\,i\, m}{\rho} \mathcal{N}_{(\mu} \eta^\perp_{\nu)k} \mathbf{Im}[\Phi^{(3/2)}(\rho, \hat{n}) \partial^k_\perp \Phi^{*(3/2)} (\rho, \hat{n})]\\
    \mp 2\, i\,m \mathcal{N}_{(\mu}\mathcal{M}_{\nu)} \mathbf{Im}[\Phi^{(3/2)}(\rho, \hat{n}) \partial_\rho\Phi^{*(3/2)} (\rho, \hat{n})] + 2 m^2 \eta_{\mu\nu} \mathbf{Re}[\Phi^{(3/2)}(\rho, \hat{n}) \Phi^{*(5/2)} (\rho, \hat{n})]
    \end{array}
\end{array}
$

From the equations of motion for a real massive scalar field, one can express $\Phi^{(5/2)}(\rho, \hat{n})$ in terms of $\Phi^{(3/2)}(\rho, \hat{n})$ in the following manner:
 \begin{equation}
     \Phi^{(5/2)}_\pm(\rho, \hat{n})=\mp\frac{i}{8 m \rho ^2} \left(3 \rho ^2 \Phi^{(3/2)} (\rho ,\hat{n})+4 \left(\partial_\perp^2\Phi^{(3/2)}
   (\rho ,\hat{n})+\rho  \left(\left(2+\rho ^2\right) \partial_\rho\Phi^{(3/2)}
   (\rho ,\hat{n})+\rho  \left(1+\rho ^2\right) \partial_\rho^2\Phi^{(3/2)}
   (\rho ,\hat{n})\right)\right)\right).
   \label{MSclr_8}
 \end{equation}
With the help of eq.\eqref{MSclr_8}, one can express $H_{\mu\nu}(\rho, \hat{n})$ in terms of $\Phi^{*(3/2)}(\rho, \hat{n})$.

Now that we have the stress-energy tensor for a real massive scalar field for $|\tau| \to \infty$, we can calculate the behavior of gravitational radiation at $\mathcal{I}^+_\pm$ and derive the classical soft graviton theorem for this particular case.

Using eq.\eqref{MSclr_7}, one gets

\begin{align}
    \Tilde{e}_{\mu\nu} (\omega, \Vec{x}) \noindent\nonumber& = \frac{e^{i\omega r}}{4 \pi r}\int dx^{\prime 4} e^{i k \cdot x^\prime} T^{(\Phi)}_{\mu\nu}\left( x^\prime \right)\\
    \noindent\nonumber&\\
    \noindent\nonumber& \simeq \frac{e^{i\omega r}}{4 \pi r} \Bigg\{\int_{\tau < -\tau_0} \frac{\tau^{\prime 3} \rho^{\prime 2}}{\sqrt{1+\rho^{\prime 2}}} d\tau^\prime d\rho^\prime d\hat{n}^\prime e^{i \omega \tau^\prime \mathcal{N}^\prime \cdot n}  \left( \frac{F_{\mu\nu}^-(\rho^\prime, \hat{n}^\prime)}{\tau^{\prime 3}} +  \frac{H_{\mu\nu}^-(\rho^\prime, \hat{n}^\prime)}{\tau^{\prime 4}}  +\mathcal{O}\bigg(\frac{1}{\tau^{\prime 5}}\bigg)\right)\\
    \noindent\nonumber&\hspace{2cm} + \int_{\tau>\tau_0} \frac{\tau^{\prime 3} \rho^{\prime 2}}{\sqrt{1+\rho^{\prime 2}}} d\tau^\prime d\rho^\prime d\hat{n}^\prime e^{i \omega \tau^\prime \mathcal{N}^\prime \cdot n}  \left( \frac{F_{\mu\nu}^+(\rho^\prime, \hat{n}^\prime)}{\tau^{\prime 3}} +  \frac{H_{\mu\nu}^+(\rho^\prime, \hat{n}^\prime)}{\tau^{\prime 4}}  +\mathcal{O}\bigg(\frac{1}{\tau^{\prime 5}}\bigg)\right) + \mathcal{O}(1)\Bigg\}\\
    \noindent\nonumber&\\
    \noindent\nonumber& = \frac{e^{i\omega r}}{4 \pi r} \Bigg\{\int_{\tau < -\tau_0} \frac{\rho^{\prime 2}}{\sqrt{1+\rho^{\prime 2}}} d\tau^\prime d\rho^\prime d\hat{n}^\prime e^{i \omega \tau^\prime \mathcal{N}^\prime \cdot n}  \left( F_{\mu\nu}^-(\rho^\prime, \hat{n}^\prime) +  \frac{H_{\mu\nu}^-(\rho^\prime, \hat{n}^\prime)}{\tau^{\prime}}  +\mathcal{O}\bigg(\frac{1}{\tau^{\prime 2}}\bigg)\right)\\
    \noindent\nonumber& \hspace{2cm}+ \int_{\tau>\tau_0} \frac{\rho^{\prime 2}}{\sqrt{1+\rho^{\prime 2}}} d\tau^\prime d\rho^\prime d\hat{n}^\prime e^{i \omega \tau^\prime \mathcal{N}^\prime \cdot n}  \left( F_{\mu\nu}^+(\rho^\prime, \hat{n}^\prime) +  \frac{H_{\mu\nu}^+(\rho^\prime, \hat{n}^\prime)}{\tau^{\prime}}  +\mathcal{O}\bigg(\frac{1}{\tau^{\prime 2}}\bigg)\right)  + \mathcal{O}(1) \Bigg\}\\
    \noindent\nonumber&\\
    \noindent\nonumber& \simeq \frac{e^{i\omega r}}{4 \pi r} \Bigg\{\frac{i}{\omega}\int \frac{\rho^{\prime 2}}{\sqrt{1+\rho^{\prime 2}}} d\rho^\prime d\hat{n}^\prime  \frac{\left(F_{\mu\nu}^+(\rho^\prime, \hat{n}^\prime) - F_{\mu\nu}^-(\rho^\prime, \hat{n}^\prime)\right)}{\mathcal{N} \cdot n}\\
    \noindent&\hspace{4cm}-\log{\omega}\int \frac{\rho^{\prime 2}}{\sqrt{1+\rho^{\prime 2}}} d\rho^\prime d\hat{n}^\prime  \left(H_{\mu\nu}^+(\rho^\prime, \hat{n}^\prime) - H_{\mu\nu}^-(\rho^\prime, \hat{n}^\prime)\right) + \mathcal{O}(1) \Bigg\}
    \label{MSclr_9}
\end{align}

We have used the following relations in the second step of eq.\eqref{MSclr_9}:
\begin{equation}
    dx^{\prime 4} \to \frac{\tau^{\prime 3} \rho^{\prime 2}}{\sqrt{1+\rho^{\prime 2}}} d\tau^\prime d\rho^\prime d\hat{n}^\prime \hspace{3mm},\hspace{3mm} k \cdot x^\prime = - \omega \tau^\prime (\sqrt{1+\rho^{\prime 2}} - \rho^\prime \vec{n} \cdot \vec{n^\prime})=-\mathcal{N}^\prime \cdot n,
    \label{MSclr_10}
\end{equation}
where $k^\mu \equiv \omega(1, \vec{n})$.

\vspace{5mm}

\hrule

\begin{center}
    \huge Appendix
\end{center}

\appendix
\section{Evaluation of few necessary integrals}

Some necessary integrals that often appear during the derivation of the classical soft graviton theorem are mentioned in the appendix. The integrations mentioned in the appendix have been previously calculated in detail by Laddha and Sen \cite{Laddha}.  

\begin{equation}
    I_{1} \equiv \frac{1}{\omega} \int_{-\infty}^{\infty} d j e^{-i \omega g(j)} f^{\prime}(j),
    \label{A.1}
\end{equation}

\begin{equation}
    I_{2} \equiv \int_{-\infty}^{\infty} d j \frac{1}{r(j)} f(j) e^{-i \omega g(j)},
    \label{A.2}
\end{equation}

where $f(j), g(j), r(j)$ are smooth functions with the property

\begin{equation}
    \begin{array}{ll}
f(j)=f_{ \pm}+\frac{k_{ \pm}}{j}, \\
g(j) \rightarrow a_{ \pm} j+b_{ \pm} \ln |j|, \\
r(j) \rightarrow c_{ \pm} j+d_{ \pm} \ln |j|, \quad \text { as } \quad j \rightarrow \pm \infty
\end{array}
\label{A.3}
\end{equation}

To evaluate integrals, one can separately estimate how the four regions: $|j| \sim 1,1 \ll\left|j\right| \ll \omega^{-1},\left| j \right| \sim \omega^{-1}$ and $|j| \gg \omega^{-1}$, contribute to the net result.

\subsection{Evaluating $I_{1}$}

$I_{1}$ can be expressed as

\begin{equation}
    I_{1}=\frac{1}{\omega} \int_{-\infty}^{\infty} d j f^{\prime}(j)+\frac{1}{\omega} \int_{-\infty}^{\infty} d j\left\{e^{-i \omega g(j)}-1\right\} f^{\prime}(j).\label{A.4}
\end{equation}

The first term in eq. \eqref{A.4} gives $\omega^{-1}\left(f_{+}-f_{-}\right)$. In the regions: $|j| \sim 1$, $|j| \sim \omega^{-1}$ and $|j|>\omega^{-1}$, the term inside the curly bracket is of order $\omega$, the second term in eq.\eqref{A.4} leads to $\mathcal{O}(1)$ contribution. In the region $1 \ll|j| \ll \omega^{-1}$, the integral can be approximated as

\begin{equation}
    -i \int_{1 \ll|j| \ll \omega^{-1}} d j g(j) f^{\prime}(j) \simeq i \int_{1 \ll|j| \ll \omega^{-1}} d j \frac{a_{ \pm} k_{ \pm}}{j} \simeq i\left(a_{+} k_{+}-a_{-} k_{-}\right) \ln \omega^{-1} + \mathcal{O}(1).
    \label{A.5}
\end{equation}

Hence, at soft limit (i.e., $\omega \to 0$),

\begin{equation}
    I_{1}=\omega^{-1}\left(f_{+}-f_{-}\right)+i\left(a_{+} k_{+}-a_{-} k_{-}\right) \ln \omega^{-1}+ \mathcal{O}(1) .
    \label{A.6}
\end{equation}

\subsection{Evaluating $I_{2}$} 

The $|j| \sim 1$, $|j| \sim \omega^{-1}$ and $|j| \gg \omega^{-1}$ regions lead to $\mathcal{O}(1)$ contributions. The contribution due to the region, $1 \ll|j| \ll \omega^{-1}$, is 

\begin{equation}
    \int_{1 \ll|j| \ll \omega^{-1}} d j \frac{f_{ \pm}}{c_{ \pm} j} \simeq \pm \frac{f_{ \pm}}{c_{ \pm}} \ln \omega^{-1}.
    \label{A.7}
\end{equation}

Hence, for $\omega \to 0$, 
\begin{equation}
    I_{2} \simeq\left(f_{+} c_{+}^{-1}-f_{-} c_{-}^{-1}\right) \ln \omega^{-1} + \mathcal{O}(1).
    \label{A.8}
\end{equation}

\bibliographystyle{unsrt}  

\begin{thebibliography}{100}

\bibitem{Sen}Saha, A.P., Sahoo, B. and Sen, A. Proof of the classical soft graviton theorem in D = 4. J. High Energ. Phys. \textbf{2020}, 153 (2020).

\bibitem{Sen_Laddha}Alok Laddha and Ashoke Sen, Classical proof of the classical soft graviton theorem in D>4, Phys. Rev. D \textbf{101}, 084011 (2020).

\bibitem{Sahoo2022}Sahoo, B., Sen, A. Classical soft graviton theorem rewritten. J. High Energ. Phys. \textbf{2022}, 77 (2022).

\bibitem{Laddha}Laddha, A., Sen, A. Logarithmic terms in the soft expansion in four dimensions. J. High Energ. Phys. \textbf{2018}, 56 (2018).

\end{thebibliography}

\end{document}